\documentclass[aaspp4,11pt,preprint,apjfonts]{emulateapj}
\slugcomment{Accepted for publication in the Astrophysical Journal}
\shorttitle{star formation in the CDFS}
\shortauthors{Damen et al.}

\newcommand{\Mstar}{\hbox{$M_*$}}
\newcommand{\Msol}{\hbox{$M_\odot$}}

\newcommand{\msol}{\hbox{$M_\odot$}}

\newcommand{\lsol}{\hbox{$L_\odot$}}

\newcommand{\um}{\hbox{$\mu$m}}
\newcommand{\elf}{\hbox{$10^{11} M_\odot$}}

\def\figa{
  \begin{figure*} 
    \includegraphics[width=\textwidth]{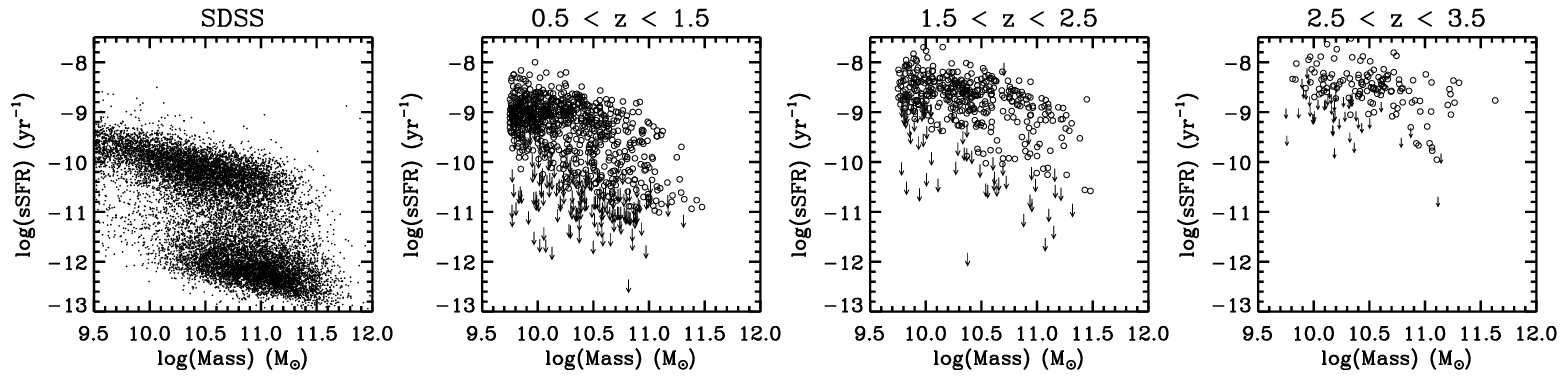}
    \caption[ssfr_mass.eps]{sSFR against mass, for the low redshift sample from the SDSS, and at mean redshifts, $z =1, 2, 3$. Open circles are the FIREWORKS detections, arrows show $2 \sigma $ upper limits of the sSFR.
\label{ssfr_m}}
  \end{figure*}
}

\def\figb{
  \begin{figure*} 
    \begin{center}
      \includegraphics[width=0.7\textwidth]{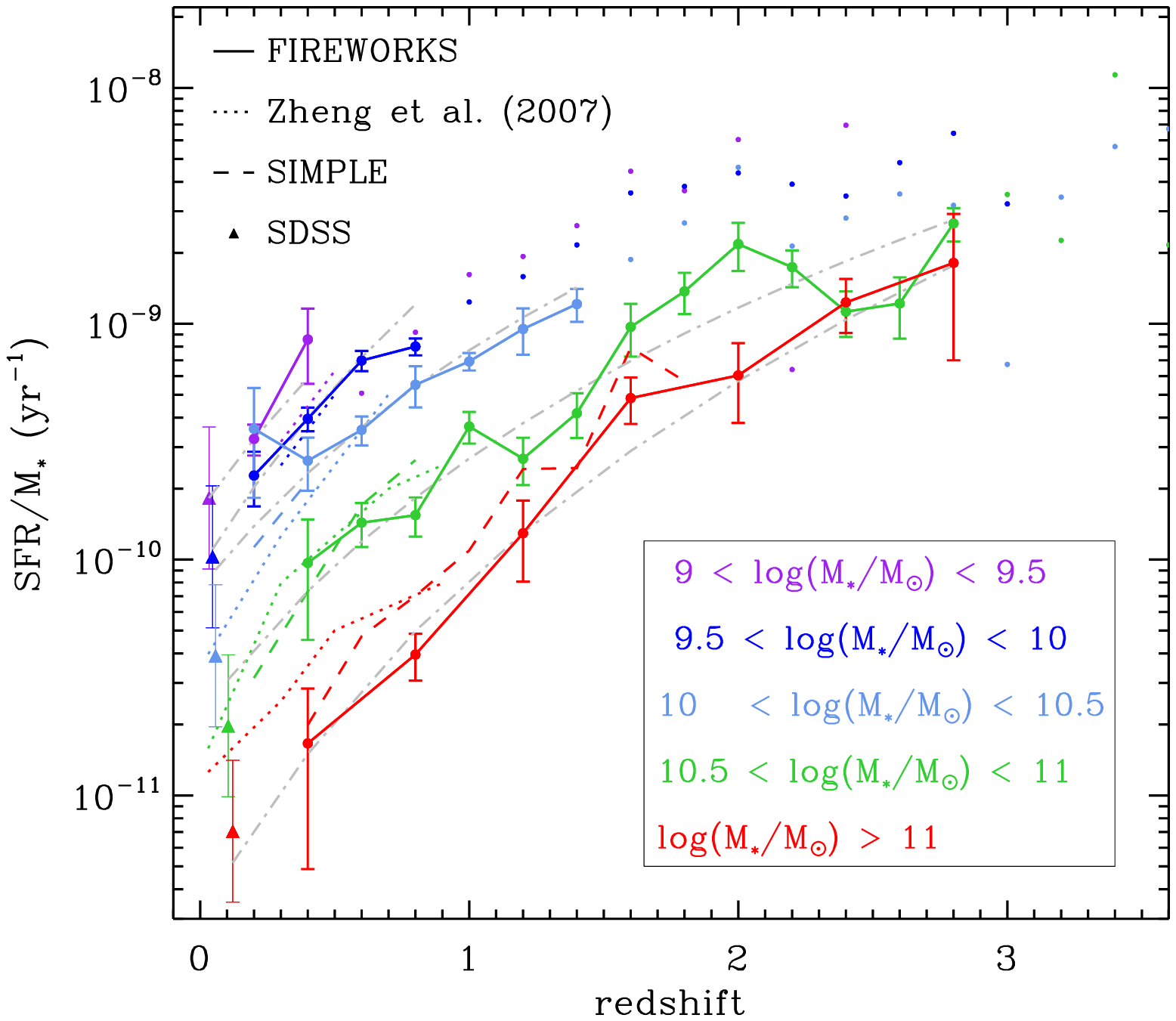}
      \caption[ssfr_z.eps]{sSFR versus redshift in different mass bins. Filled, connected circles are the FIREWORKS results, dots show where mass incompleteness starts to play a role. The error bars represent bootstrap errors. The dashed and dotted lines represent results from Damen et al. (2009) and Zheng et al. (2007), respectively. SDSS data were used to include a local data point ({\it triangles}). Linear fits to the FIREWORKS results are displayed using gray dash-dotted lines. The fits show that the SSFR increases with $z$ at a rate that appears independent of mass.
        \label{ssfr}}
      \end{center}
    \end{figure*}
}

\def\figc{
  \begin{figure*} 
    \includegraphics[width=\textwidth]{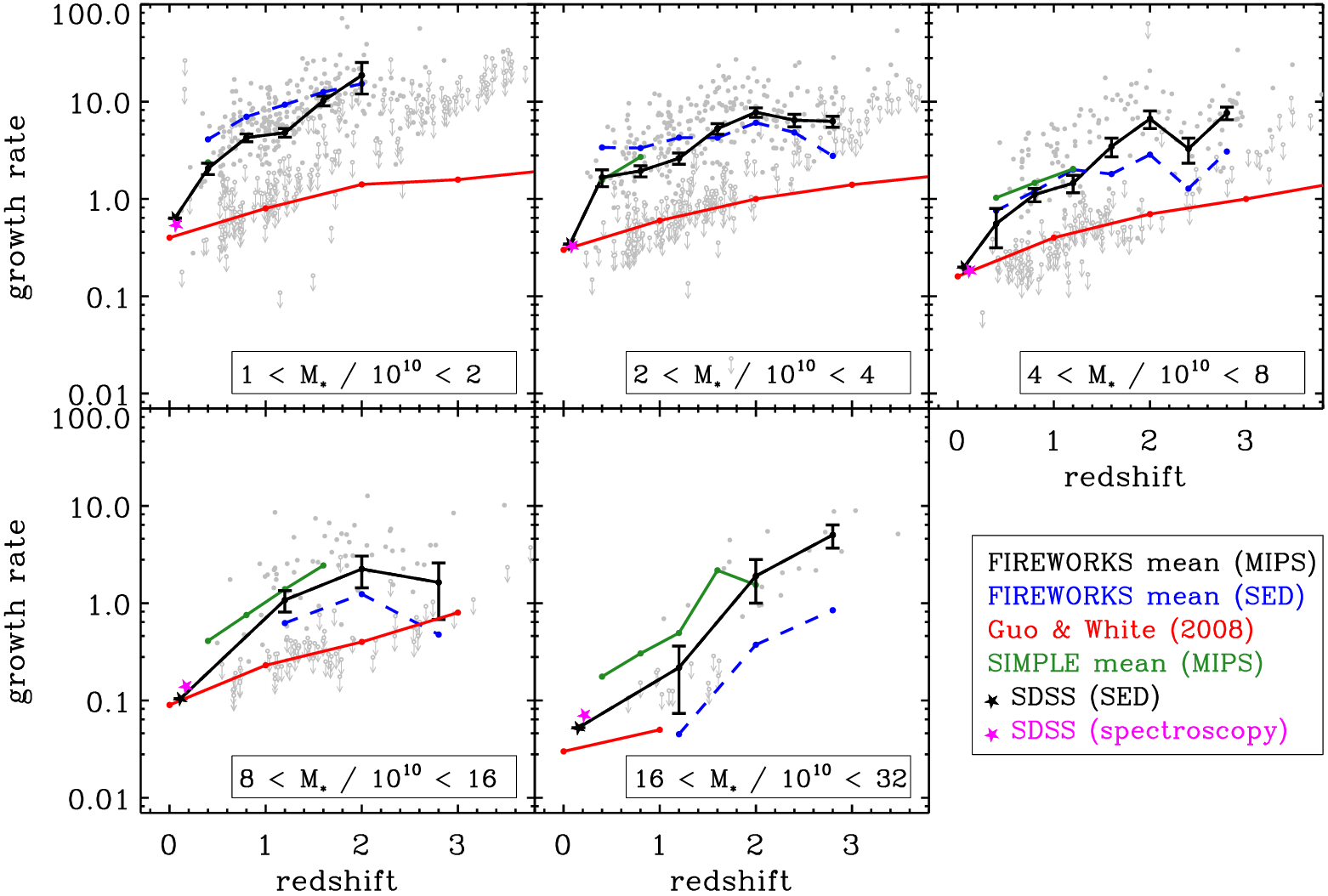}
    \caption[guo_comp.eps]{Dimensionless growth rate through star formation as a function of redshift. The different panels represent different mass bins in units of $10^{10}$ \msol. The gray dots are the FIREWORKS values, the black line is their mean. Open circles with arrows denote upper limits set at the $2 \sigma$ level. Overplotted in red are the results from Guo \& White (2008), based on the Millennium Simulation. Additional observational measurements are added, based on SFRs determined from SED fitting ({\it blue line}) and the results from SIMPLE ({\it green line}, Damen et al. 2009). To represent the local universe, both a photometric ({\it black star}) and a spectroscopic ({\it magenta star}) measurement from SDSS were included. Note the large offset between the model and the observations between redshift z=0 and z=1. The model and observations both show an increase of the growth rate with redshift, but for the model this trend is more gentle. 
 \label{guo}}
  \end{figure*}
}

\def\figd{
  \begin{figure} 
    \includegraphics[width=0.5\textwidth]{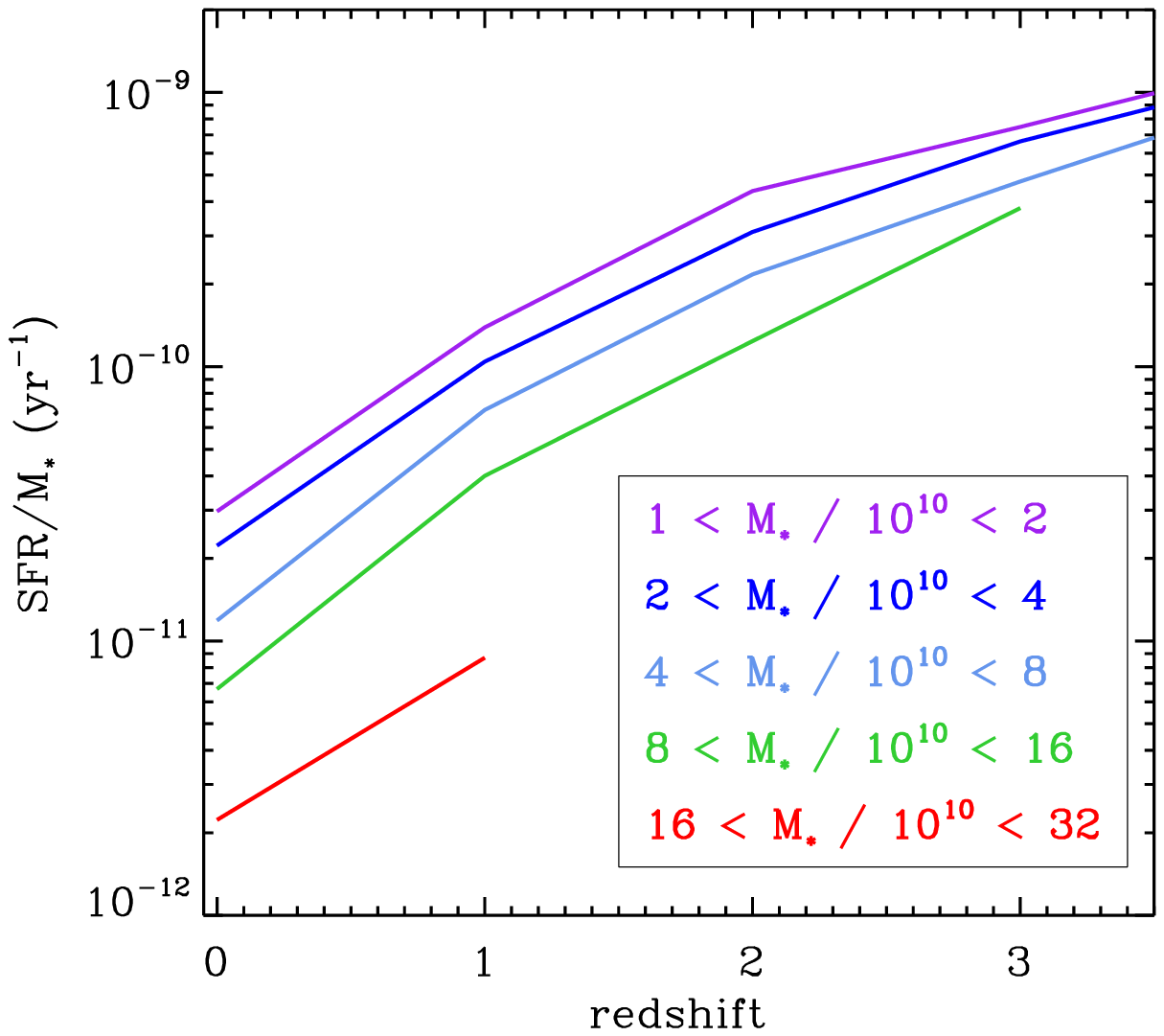}
    \caption{sSFR versus redshift  for the Guo \& White (2008) results. Different colors represent the mass bins used in Figure \ref{guo}. The slope of the evolution of sSFR with redshift is nearly independent of mass, which agrees with the FIREWORKS results.
  \label{guo_ssfr}}
  \end{figure}
}

\def\fige{
  \begin{figure}
    \includegraphics[width=0.5\textwidth]{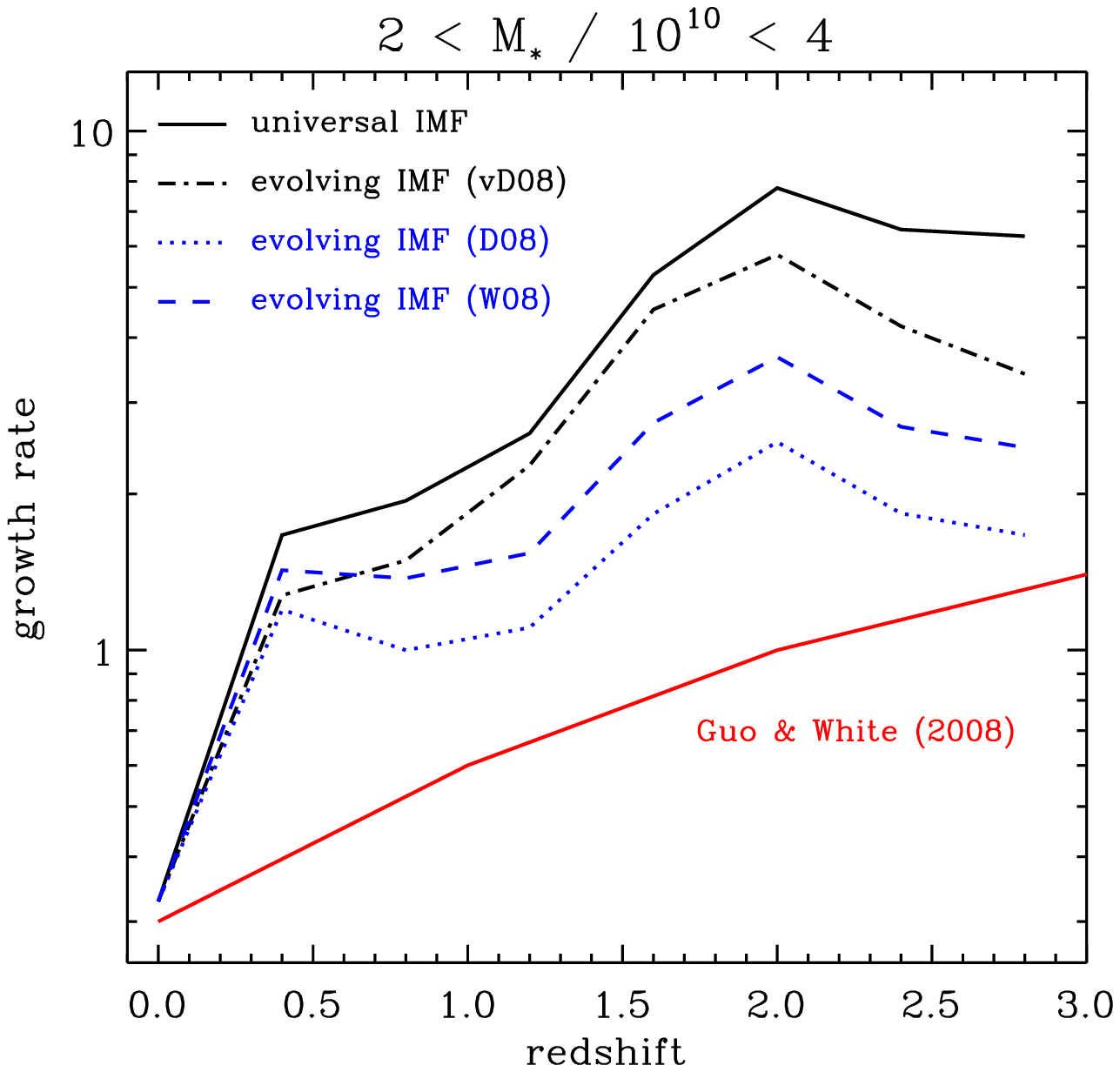}
    \caption[corr.eps]{Dimensionless growth rate as a function of redshift for galaxies with masses ranging from $2 \cdot 10^{10} \msol$ to $4\cdot 10^{10} \msol$. The black solid line shows the FIREWORKS result when a Kroupa (2001) IMF is applied. The dashed-dotted lines, dashed and dotted lines show the FIREWORKS growth rate based on an evolving IMF, according to the parametrization of van Dokkum (2008; vD08), Dav\'e (2008; D08), and Wilkins et al. (2008; W08), respectively. The corrected values based on the IMFs of D08 and W08 ({\it in blue}) are lower limits, since they do not include the effect an evolving IMF has on the stellar mass. Introducing a time-dependent IMF decreases the discrepancy between the observations and the simulated results from Guo \& White (2008; {\it red solid line}), but it does not completely resolve it. In particular, the steep increase in observed growth rate at low redshift (z = 0 - 1) is still evident.
  \label{IMF_corr}}
  \end{figure}
}

\begin{document}

\title{Star formation in the CDFS: observations confront simulations}

\author{Maaike Damen\altaffilmark{1}, 
Natascha M. F\"orster Schreiber\altaffilmark{2}, 
Marijn Franx \altaffilmark{1}, 
Ivo Labb\'e \altaffilmark{3}, 
Sune Toft\altaffilmark{4},
Pieter G. van Dokkum \altaffilmark{5}, 
Stijn Wuyts \altaffilmark{6}}

\email{damen@strw.leidenuniv.nl}

\altaffiltext{1}{Leiden Observatory, Leiden University, PO Box 9513,
  2300 RA Leiden, The Netherlands}
\altaffiltext{2}{Max-Planck-Institut f\"ur extraterrestrische Physik, Giessenbachstrasse, 
D-85748, Garching, Germany}
\altaffiltext{3}{Carnegie Observatories, 813 Santa Barbara Street, Pasadena, CA 91101; Hubble Fellow}
\altaffiltext{4}{Dark Cosmology Centre, University of Copenhagen, Blegdamsvej 172100, Copenhagen, Danmark}
\altaffiltext{5}{Department of Astronomy, Yale University, New Haven, CT, 06520}
\altaffiltext{6}{Harvard-Smithsonian Center for Astrophysics, 60 Garden Street, Cambridge, MA 02138}

\begin{abstract} 
We investigate the star formation history of the universe using FIREWORKS, a multiwavelength survey of the CDFS. We study  the evolution of the specific 
star formation rate (sSFR) with redshift in different mass bins from z = 0 to $z \sim 3$. We find that the sSFR increases with redshift for all masses. 
The logarithmic increase of the sSFR with redshift is nearly independent of mass, but this cannot yet be verified  at the lowest-mass bins at $z > 0.8$, 
due to incompleteness. We convert the sSFRs to a dimensionless growth rate to facilitate a comparison with a semi-analytic galaxy formation model that was implemented on the Millennium Simulation. The model predicts that the growth rates and sSFRs increase similarly with redshift for all masses, consistent with the observations. However, we find that for all masses, the inferred observed growth rates increase more rapidly with redshift than the model predictions. We discuss several possible causes for this discrepancy, ranging from field-to-field variance, conversions to SFR, and shape of the IMF. We find that none of these can solve the discrepancy completely. We conclude that the models need to be adapted to produce the steep increase in growth rate between redshift $z=0$ and $z=1$. 
\end{abstract}

\keywords{galaxies: evolution --- galaxies: formation --- galaxies: 
  high-redshift}

\section{Introduction}
To understand galaxy formation and evolution, it is essential to have a clear idea of how galaxies assemble their mass. Multiwavelength galaxy surveys of the high-redshift universe provide estimates of masses and star formation  rates out to $z \sim 6$ (Lilly et al. 1996, Madau et al. 1996, and the compilations by Hopkins (2004) and Hopkins \& Beacom (2006)). Recent studies of the evolution of the SFR per unit mass (specific SFR, sSFR) have shown that this quantity is an increasing function with redshift out to $z \sim 2$ and that, at a given redshift, the most massive galaxies typically have the lowest specific SFRs (Juneau et al. 2005, Bauer et al. 2005, P\'erez-Gonz\'alez et al. 2005; 2008, Caputi et al. 2006, Papovich et al. 2006, Reddy et al. 2006, Noeske et al. 2007, Martin et al. 2007, Zheng et al. 2007, Damen et al. 2009). Many of these studies support the idea of downsizing, where the locus of active star formation shifts from massive galaxies to less massive galaxies with time (Cowie et al. 1996).\\
It is interesting to compare these observations to model predictions. Using semi-analytical modeling (SAM) techniques, the formation of galaxies can be simulated within the standard cold dark matter ($\Lambda$CDM) cosmogony. SAMs have been relatively successful in reproducing numerous systematic properties of the observed local population at $z < $0.4, such as number densities, luminosity functions, mass functions, and SFRs (e.g. Madau, Pozzetti \& Dickinson 1998, Cole et al. 2000, Fontana et al. 2006, Fontanot et al. 2009). Since the recipes in the SAMs are arguably tuned to obtain such a good match, it is an interesting exercise to extend the comparison with observations to higher redshift, leaving the parameters of the SAMs to their locally tuned values. Several studies make the comparison for the high-redshift universe (Bower et al. 2006; Kitzbichler \& White 2007; Marchesini \& van Dokkum 2007; Dav\'e 2008; Genzel et al. 2008; Guo \& White in preparation, Elbaz et al. 2007, Daddi et al. 2007a, Santini et al. 2009). \\
\figa
Bower et al. (2006) find their model to be in good agreement with the observed mass function out to $z \sim 2$. However, their models overpredict the star formation density by $\sim$ 20\% at low redshift ($z < 0.4$), whereas at $z \sim 1$ they underpredict the observations by a similar factor. Refinements to the code with respect to heating by active galactic nuclei (AGN; Bower et al. 2008) have not yet helped to reduce the relatively small discrepancies. A similar effect, where the models underpredict star formation observed at high redshift, is also seen in hydrodynamical simulations (both for multiwavelength (Dav\'e 2008) and spectroscopic (Genzel et al. 2008) observations. Kitzbichler \& White (2007) find an abundance of massive galaxies that overpredicts the observed abundance by a factor of $\sim$2 at $z = 2$. Studies by Elbaz et al. (2007), Daddi et al. (2007a), and Santini et al. (2009), reveal how SAMs underpredict the observed star formation rate by a factor 2-5 out to redshift $z=2$.\\
In this paper we explore the evolution of the star formation history of galaxies from redshift $z \sim 3$ to 0, using FIREWORKS, a $K_S$-band selected multiwavelength catalog of the CDFS, that includes deep MIPS data to allow derivation of SFRs. Galaxy growth can be conveniently quantified using the dimensionless growth rate, defined as sSFR $* t_H(z)$, where $t_H(z)$ is the Hubble time (Guo \& White 2008). It provides a direct and quantitative constraint on the models. We first compare our results to other observational studies and then, using the growth rate, make a comparison with a SAM built on top of the Millennium Simulation. Throughout this paper we assume a $\Lambda$CDM cosmology with $\Omega_{\rm m}=0.3$, $\Omega_{\rm  \Lambda}=0.7$, and $H_{\rm 0}=70$~km s$^{-1}$ Mpc$^{-1}$. All magnitudes are given in the AB photometric system. Stellar masses are determined assuming a Kroupa (2001) initial mass function (IMF).

\section{Data}\label{dat}
\subsection{Observations and sample selection}\label{sam}
We use the FIREWORKS catalog for the GOODS-CDFS, generated by Wuyts et al. (2008). This catalog is based on high quality photometry ranging from the NUV to MIR. Wuyts et al. (2008) included deep space based optical imaging with HST using ACS (Giavalisco, Steidel \& Macchetto 1996). They complemented this with optical imaging obtained as part of the COMBO-17 (Wolf et al. 2003) and ESO DPS (Arnouts et al. 2007) surveys, in the re-reduced form of the GaBoDs consortium (Erben et al. 2005; Hildebrandt er al. 2006). NIR imaging was obtained with VLT/ISAAC (Vandame 2002, Vandame et al. in preparation). For mid-IR wavelength coverage they used deep Spitzer imaging (IRAC and MIPS) from Dickinson et al. (in preparation).\\
A $K_s$-selected catalog was constructed following the procedures of Labb\'e et al. (2003) and contains the following bands:  $U_{38}BVRI$ (WFI), $B_{435}V_{606}i_{775}z_{850}$ (ACS), $JHK_s$ (ISAAC), 3.6-8.0 \um \, (IRAC) and 24 \um \, (MIPS). It has a 5 sigma depth in $K_s$ of $\sim$24.3 and a total area of 138 arcmin$^2$. For details on observations, source detection and astrometry we refer to Wuyts et al. (2008). Using the CDFS X-ray catalog of Giacconi et al. (2002), we excluded all X-ray detected sources from the sample as they are likely AGN. We further restricted the selection to sources with a signal-to-noise higher than 10 in the $K_s$-band, which results in a total sample size of 5,274 sources.

\subsection{Redshifts, masses and star formation rates}
Wuyts et al. (2008) derived photometric redshifts using EAZY (Brammer et al. 2008) which are in good agreement with the available spectroscopic redshifts. The normalized median absolute deviation (NMAD) of $(z_{\rm phot} - z_{\rm spec})/(1+z_{\rm spec})$  is 0.031 over the whole redshift range and NMAD = 0.071 at $z > 1.5$ (Wuyts et al. 2008). 
To obtain stellar masses, ages, dust attenuation and star formation rates, stellar population models were fitted to the spectral energy distribution (SED) out to 8 \um. The SED modeling follows standard procedures, and is described by F\"orster Schreiber et al. (in preparation; see also e.g., F\"orster Schreiber et al. 2004). Bruzual \& Charlot (2003) models were used with solar metallicity, a Salpeter IMF and a Calzetti reddening law. Masses and SFRs were converted to a Kroupa IMF by subtracting 0.2 dex. The fits were performed with the HYPERZ program (Bolzonella, Miralles \& Pell\'o 2000), fixing the redshift to the photometric redshift derived with EAZY, or the spectroscopic redshift when available. Three star formation histories were fit and the best fitting model was used. The three star formation histories were: a single stellar population without dust, a constant star formation history and an exponentially declining star formation history with a characteristic timescale of 300 Myr, the latter two with varying amounts of dust ($0 < A_v < 4$). A full description of the SED fitting procedure and extensive tests on the outcome can be found in F\"orster Schreiber et al. (in preparation). Rest-frame luminosities were derived by interpolating between observed bands using the best-fit templates as a guide (see Rudnick et al. (2003) for a detailed description of this technique and Taylor et al. (2009) for the IDL implementation of the algorithm, dubbed 'InterRest'\footnote{http://www.strw.leidenuniv.nl/$\sim$ent/InterRest}).\\
In addition to the SFRs derived from SED fitting, SFRs were determined independently using a combination of rest-frame UV and IR emission. Between $1 < z < 3$, MIPS 24 \um \, traces the rest-frame 6-12 \um\, luminosity, which correlates broadly with the total IR luminosity. Using a wide range of templates from Dale \& Helou (2002), bolometric luminosities were determined, adopting the mean of log$(L_{IR})$, following Wuyts et al. (2008) and Labb\'e et al. (in preparation). In other words, we apply a linear conversion from observed 24 \um\, flux to $L_{IR}$, using a coefficient which depends also on redshift. The SFR was determined assuming 
\begin{equation}\label{eqn:ir_sfr}
\Psi / \msol \,\,\mathrm{yr}^{-1} = 1.1 \times 10^{-10} \times
(L_\mathrm{IR} + 3.3\,\, L_{2800}) / \lsol,
\end{equation}
based on the conversion by Bell et al. (2003) converted to a Kroupa IMF (see also Wuyts et al. (2008), Labb\'e et al. (in preparation)). Unless stated otherwise, we use the MIPS derived SFRs in our analysis. The SFR determined from the SED fitting is used for comparison in later sections.\\
We include data from the Sloan Digital Sky Survey (SDSS) to extend our redshift coverage to $z=0$. SDSS masses were determined by Kauffmann et al. (2003) using spectra. Brinchmann et al. (2004) derived SFRs from emission lines. For details on the derivation of the masses and SFRs in the SDSS we refer to their papers.\\
Figure\, \ref{ssfr_m} shows the specific star formation rate with mass in four different redshift regimes ($ z = 0, 1, 2, 3$). The first panel shows SDSS results, the others are based on the FIREWORKS data. Although the spread in sSFR is high at a given mass (see also Franx et al. 2008), two trends are clear. Galaxies of higher mass typically have lower sSFRs and at fixed mass the sSFR rises with redshift. Both trends have been observed before and confirm earlier results (Bauer et al. 2005, Zheng et al. 2007, P\'erez-Gonz\'alez et al. 2008, Damen et al. 2009). We will discuss the evolution of the sSFR further in Section \ref{ssfr_z_sect}.

\figb

\begin{deluxetable}{lc}
  \tablecolumns{2}
    \tablewidth{0pc}
    \tablecaption{75\% completeness limits\label{comp.tab}
    }
    \tablehead{
      \colhead{log$(\Mstar/\Msol)$} & \colhead{$z$}
    }
    \startdata
    9.0 -9.5     & 0.5\\
    9.5 -10.0    & 0.9\\
    10.0 -10.5   & 1.5\\
    10.5 -11.0   & 2.9\\
    $>$11.0    & 2.9
    \enddata
\end{deluxetable}

\subsection{Mass completeness}
We determine the mass completeness limit using our $K_s$-band selection limit. We scale the masses of sources down to flux selection limit of 10 $\sigma$ ($M_{scaled} = M \times 10 / [S/N]_{K_S}$) and determine the mass limit to which we can detect 75\% of the sources at $10\, \sigma$ in a narrow redshift bin. We find that our sample is 75 \% complete for masses higher than \elf \,out to $z \sim 3$. The completeness limits are listed in Table~\ref{comp.tab}. In the remainder of the paper, all references to completeness are based on the 75\% completeness limits.\\

\section{The evolution of the specific star formation rate}\label{ssfr_z_sect}
In a previous paper (Damen et al. 2009) we studied how the SFR changes with mass and redshift out to $z \sim 2$. We used the SIMPLE survey, which is based on NUV to NIR observations of the E-CDFS. With an area of 0.5 x 0.5 deg, the E-CDFS is $\sim$6.5 times larger than the much deeper FIREWORKS area. Here we extend this analysis to higher redshift and lower masses, using the FIREWORKS survey of the CDFS. We determine the average SFR in five different mass bins and investigate the evolution of the specific SFR with redshift. For sources with no significant MIPS flux, we did not use upper limits to determine the average, but included the measured fluxes.
Our results are shown in Figure~\ref{ssfr}. Different colors represent different mass bins and dots denote where we suffer from incompleteness. Error bars represent the bootstrapped 68\% confidence levels on the average sSFRs. Local SDSS data points (triangles) were added to the figure and lie on the same trend as our results. We also compare our results to previous similar studies. Dotted lines show the results of Zheng et al. (2007) who used the COMBO-17 survey. Dashed lines represent the work of Damen et al. (2009), based on the SIMPLE survey.\\
Those papers showed that sSFRs of massive galaxies are typically smaller than those of less massive galaxies and that the increasing trend of sSFR with redshift has a similar slope for all masses. For the lowest to average mass bins ($9.0 <\, $log$(M/\Msol) < 11$) we confirm results from Zheng et al. (2007) and Damen et al. (2009) and extend them out to higher redshifts. However, the sSFRs in the highest mass bin ($M_* > 10^{11} \Msol$), seem to be lower than both the results from SIMPLE and Zheng et al. 2007. This can be a matter of low number statistics, since the average number of galaxies per redshift bin for these masses is $\sim 8$. \\
\begin{deluxetable}{lc}
\tablecolumns{2}
\tablewidth{0pc}
\tablecaption{slope of SSFR-$z$ relation \label{n.tab}
}
\tablehead{
\colhead{log$(\Mstar/\Msol)$} & \colhead{$n$}
}
\startdata
9.5 -10   & 4.40$\pm$ 0.31\\
10  -10.5 & 3.36$\pm $0.87\\
10.5-11   & 3.63$\pm$ 0.33\\
  $>$11  & 4.78$\pm$ 0.37
\enddata
\end{deluxetable}
To quantify the slopes of the trends in each mass bin, we fitted the sSFR with $(1+z)^n$ over the redshift range where it is complete with respect to mass. We used a bootstrapping technique to determine errors on the fits. The resulting values for the steepness of the slope, $n$, are listed in Table~\ref{n.tab}, the fits themselves are shown in Figure \ref{ssfr} by dash-dotted lines, in gray. \\
It is interesting to regard these values in the context of downsizing. We can see that at fixed redshift, the sSFR decreases with increasing mass. This indicates that the most massive galaxies have already formed the bulk of their stars and that active star formation has shifted to galaxies that are less massive. However, we do not see a strong mass dependence in the decline of the sSFR with time. This confirms results from Zheng et al. (2007) and Damen et al. (2009), whereas it seems to disagree with the results of Juneau et al (2005), who find that, out to $z=1.5$, the sSFRs of more massive galaxies decrease faster in time than those of the lowest-mass galaxies. Taking into account the error bars in their Figure 3, this disagreement could be subtle. \\
Admittedly, the redshift range over which we determined $n$ is small for the low-mass bins (see Figure~\ref{ssfr}). In addition, the highest mass bin (log$(\Mstar/\Msol) > 11$ suffers from low number statistics. We verified whether our results would change with a larger number of sources by including the SIMPLE data, and find that the trend in Figure~\ref{ssfr} at the high-mass end remains the same. Including the SIMPLE data does not alter the results in a way that evolution in the slope can be confirmed. To be more conclusive, these fits need to be determined over similar redshift ranges for all mass bins. Deeper observations, that solve the incompleteness issues, are needed to accurately make a statement about the possible evolution of $n$ with mass.

\figc
\section{Comparison with model predictions}\label{modcomp}
Next, we compare our results with model predictions. We use the work of Guo \& White (2008), who have used data from the Millennium Simulation to study the growth of galaxies through mergers and star formation in the semi-analytic models of de Lucia \& Blaizot (2007)\footnote{Please note that the de Lucia \& Blaizot (2007) model assumes a Chabrier IMF. This will not significantly affect the comparison, since it greatly resembles the Kroupa IMF that we apply.}.\\
In a qualitative way, the results of Guo \& White (2008) concerning star formation are consistent with the observed trends in Figure~\ref{ssfr}. They find that the growth rates through star formation increase rapidly with redshift, for all stellar masses. In addition, they find that specific growth rates through star formation are smaller in high-mass than in low-mass galaxies at all redshifts. Although qualitatively their model predictions agree with our results, quantitatively this is not the case. We determine dimensionless growth rates for our sample similar to Guo \& White (2008). The growth rate is defined as $GR_{sf} = SFR/M_* \cdot t(z)$: the sSFR multiplied with the age of the universe at the redshift of observation. 
We determine mean values of this growth rate at all redshifts, in the same mass bins Guo \& White (2008) use. The results can be seen in Figure~\ref{guo}. The gray dots and black line represent the FIREWORKS growth rates and their binned mean. The results from Guo \& White (2008) are shown in red. For comparison with the local universe we used SFRs and masses from the SDSS to determine growth rates at $z = 0$. We included both estimates from SED fitting (J. Brinchmann, private communication, black star) and from emission lines (Brinchmann et al. 2004, magenta star). The SAMs are tuned to fit the local universe and, except for the highest mass bin, the agreement is excellent. \\
Although the growth rates agree at redshift $z=0$, at higher redshifts the observed growth rates increase much more rapidly than the model growth rates. The overall offset between redshifts $z = 1$ and $z = 2$ is a factor of $\sim$6.3 for all masses. Per mass bin, we find offsets by factors of 4.0 ($8 \cdot 10^{10} \msol -16\cdot 10^{10} \msol$), 6.2 ($4 \cdot 10^{10} \msol -8\cdot 10^{10} \msol$), 6.0 ($2 \cdot 10^{10} \msol -4\cdot 10^{10} \msol$) and 9.1 ($10^{10} \msol - 2 \cdot 10^{10} \msol$) at redshifts $z=1-2$. This agrees well with recent results in the literature. Elbaz et al. (2007) and Daddi et al. (2007a) found an excess in observed sSFRs with respect to those predicted by the Millennium Simulation of a factor of $\sim$ 2.5 ($z \sim $1) and $\sim$ 4 ($z \sim $2), respectively, for galaxies of $3\cdot 10^{10} \msol$. Recent work by Santini et al. (2009) shows a similar trend where the Millennium Simulation undepredicts star formation activity by a factor 3-5. Though both these studies and our own are based on observations of the same field, there are a differences in sample selection, determination of SFRs and photometric redshifts. Given the wide variety in methods and samples used, it is remarkable that the observations converge to consistent results, namely high sSFRs with respect to model predictions.\\
This discrepancy could have its origin in the derivation of our SFRs based on the MIPS 24 \um \, fluxes. To test this we include the growth rates derived from the SFRs from our SED fits (blue lines). The growth rates derived from the bolometric fluxes and the SED fits are generally consistent with each other, except at the highest masses. This result is remarkable and it argues against the hypothesis that a simple conversion error in the calculation of the SFR causes the problem. It is interesting to note that the SED fits generally give lower sSFRs for the highest mass galaxies than the sSFRs based on UV + 24 \um. AGN may be partially responsible (Daddi et al. 2007b). We will return to the effect of errors in the SFR determination in the next section.\\
A different cause could be that the CDFS suffers from cosmic variance. 
We added in the results from the SIMPLE survey, which is based on observations on the E-CDFS and has $\sim$6.5 times more area (green lines). Both fields show similar results. As an additional exercise, we redetermined the SIMPLE growth rates excluding the CDFS and for the CDFS only, to see whether the different areas show substantial differences in growth rates. We find that for all masses and at all redshifts, the growth rates for both areas are consistent with each other within $1 \sigma$. Field to field variation, therefore, does not seem to cause a significant error.\\
It is interesting to note, that although the steepness of the observed slope is not reproduced by the model, the slope is the same for galaxies of low and high mass. To illustrate this, we used the model growth rates to derive sSFRs for the Guo \& White (2008) data. The results are shown in Figure \ref{guo_ssfr}. Different colors represent the same mass bins used in Figure \ref{guo}. Both for the model and observations, the rate of increase of sSFR does not seem to be a strong function of mass.

\figd
\fige

\section{Discussion}
We use the FIREWORKS survey to investigate the star formation history in the CDFS. We find that the sSFR is an increasing function with redshift for all masses. The slope of the trend is similar for galaxies of all masses, although some uncertainties remain due to incompleteness at the low-mass end and low number statistics in the highest mass bin ( $>$ \elf). These findings are in agreement with results from previous studies (Martin et al. 2007, Zheng et al. 2007, P\'erez-Gonz\'alez et al. 2008, Damen et al. 2009). \\
We also compare our results to model predictions of Guo \& White (2008), who use the galaxy formation model of de Lucia \& Blaizot (2007) based on the Millennium Simulation and find that for both the model and the observations the growth rate increases with redshift for all masses.
However, the model fails to reproduce the steep rise that is observed in the data, particularly between $z = 0$ and $z = 1 $, a redshift range where observations are most reliable. The overall value of the growth rate is higher in the observations than in the models, by a factor of $\sim$6.3 on average. This discrepancy between the model and the data is caused by problems with the observations, the model, or both. \\
First of all, field to field variation may play a role. We have already discussed in Section~\ref{modcomp} that this is unlikely to be a large effect. An additional factor of uncertainty resides in our estimate of the SFR based on MIPS fluxes. AGNs would contribute to the 24 \um \,emission and generate overestimated SFRs. Although we removed all X-ray detections from our sample, dust-obscured AGNs that show strong 24 \um \, flux can not be identified. Furthermore, systematic uncertainties in the conversion from MIPS fluxes to total IR luminosities are of the order of a factor 3 at $z \sim 2$ (Labb\'e et al. in preparation). In addition, we can not be sure that the local IR templates on which we base our conversion still hold in the $ z > 1$ universe. Marcillac et al. (2006) suggest that local templates are still accurate out to $z \sim 1$. Papovich et al. (2007) argue that at $z \sim 2$, local templates overpredict the total IR luminosity by a factor of a few. However, their local templates include a luminosity dependent conversion. In contrast, we apply a single template (without luminosity dependence), which results in smaller systematic offsets. In addition, we found good agreement between the average SFRs based on MIPS flux and SED modeling for the low- to intermediate-mass galaxies. It is interesting to compare and investigate other measurements of the SFR. SFRs based on H$_{\alpha}$, for instance, agree quite well with SFRs based on SED modeling (F\"orster Schreiber submitted). Furthermore, growth rates based on H${_\alpha}$ show significant disagreement with model predictions (Genzel et al. 2008), similar to what we find. Concluding, it is unlikely that the full discrepancy is caused by the systematic uncertainties in deriving SFRs. \\
A third reason why our SFRs are large, could be the assumption of a non-evolving IMF. Recently, there have been several authors who suggest that the IMF is not constant, but instead evolves with redshift. Van Dokkum (2008; vD08) found that reconciling the evolution of color and mass-light-ratio of massive, early-type cluster galaxies favors a top-heavy IMF. Dav\'e (2008; D08) argues that an evolving IMF brings the observed evolution in the relationship between SFR and stellar mass in better agreement with model predictions. Finally, Wilkins et al. (2008; W08) claim that the assumption of a simple model for an evolving IMF significantly reduces the discrepancy between the integrated SFH and stellar mass density measurements.\\
We used the effect an evolving IMF would have on SFRs (Figure\,3 from W08, Figure\,3 from D08) and SSFRs (Figure\,12a and Figure\,14b from vD08) to redetermine our growth rates and show the results in Figure\,\ref{IMF_corr}\footnote{The model predictions were kept to the values based on the non-evolving Chabrier IMF}. It is important to note that for the W08 and D08 growth rates (shown in blue), we only included the effect an evolving IMF has on SFR, and not on mass. For this reason the blue curves should be regarded as lower limits. The overall factor with which the growth rate decreases when an evolving IMF is applied is $\sim$1.3 for van Dokkum (2008), $\lesssim$1.8 for Wilkins et al. (2008) and $\lesssim$ 2.4 for Dav\'e (2008). Although this helps reducing the discrepancy between the instantaneous SFR and SFR based on the mass function (W08), the steep increase between z = 0 and z = 1 is still intact.\\
Finally, the use of photometric redshifts introduces some uncertainty to the SFR estimates. We note, however, that the growth rates of SIMPLE (green line in Figure~\ref{guo}) are based on the photometric redshifts of COMBO-17, which are highly accurate out to $z\sim0.8$. Since the SIMPLE results show the same strong slope as the FIREWORKS growth rates, we do not expect photometric redshift errors to be a significant cause for the discrepancy between the model and observations.\\
None of the effects we discussed above can provide a clear-cut solution to explain the difference between the model and observations, particularly the observed steep rise in growth rate between z=0 and z=1. The discrepancy could be caused by a flaw in the models, e.g. concerning for instance the gas supply to the galaxy. The current way stars are formed in SAMs is by heating gas to the virial temperature after which it gradually cools to the temperature where star formation can commence. Dekel et al. (2009) show that this process is too slow to explain the fact that massive galaxies have already formed most of their stars at high redshift. They introduce a new way of star formation, where cold gas enters the dark matter halo through filaments and can start forming stars immediately. It would be interesting to investigate how this implementation of star formation would alter model predictions and whether it would be able to reproduce the steep rise of the growth rate at low redshifts. \\
Despite the differences we identified between our observations and the model values of Guo \& White (2008), we also found an interesting agreement. It is remarkable that both observations and models find that the increase in sSFR is the same for low- and high-mass galaxies. There is no evidence for a ``break'' redshift at any of the masses studied here. Any modification in model recipes would have to maintain this mass independence.\\ \\

\acknowledgments
We thank the referee for useful suggestions. This research was supported by grants from the Netherlands Foundation
for Research (NWO), and the Leids Kerkhoven-Bosscha Fonds. Support from National
Science Foundation grant NSF CAREER AST-0449678 is gratefully acknowledged.
N.~M.~F.~S. acknowledges support by the Minerva Program of the Max-Planck-Gesellschaft, S.~W. acknowledges support from the W.~M. Keck Foundation, and S.~T. acknowledges support from the Lundbeck foundation.

\end{document}